\documentclass[aip,cha,preprint,numerical]{revtex4-1}
\usepackage{amsmath}
\usepackage{amssymb}
\usepackage{graphicx}
\usepackage{dcolumn}
\usepackage{bm}
\usepackage{natbib}		
\usepackage{color}

\newcommand{\ee}{\mathrm{e}}
\newcommand{\ii}{\mathrm{i}}
\newcommand{\dd}{\mathrm{d}}

\newcommand{\pd}[2]{\displaystyle\frac{\partial #1}{\partial #2}}

\newcommand{\der}[2]{\displaystyle\frac{\dd #1}{\dd #2}}

\newcommand {\e} {\varepsilon}
\newcommand {\vp} {\varphi}
\def\const{\mbox{const}}
\def\w{\omega}
\def\W{\Omega}
\newcommand{\avr}[1]{\langle #1 \rangle}

\begin{document}
\title{Dynamics of globally coupled oscillators: 
progress and perspectives}

\author{Arkady Pikovsky}
\affiliation{Institute of Physics and Astronomy, University of Potsdam, 
Karl-Liebknecht-Str. 24/25, 14476 Potsdam-Golm, Germany}
\author{Michael Rosenblum}
\affiliation{Institute of Physics and Astronomy, University of Potsdam, 
Karl-Liebknecht-Str. 24/25, 14476 Potsdam-Golm, Germany}

\date{\today}

\begin{abstract}
In this paper we discuss recent progress in  
research of ensembles of mean field coupled oscillators.  
Without an ambition to present a comprehensive review, 
we outline most interesting from our viewpoint results and
surprises, as well as interrelations between different approaches.
\end{abstract}

\maketitle

\begin{quotation}
Studies of large systems of interacting oscillatory elements is a popular and 
extensively developing branch of nonlinear science. The number of publications
on the subject grows rapidly, with many crucial contributions published 
in the Chaos journal. 
In addition to purely academic 
interest, this research finds promising applications in various fields. 
Representative examples are understanding of pedestrian synchrony on footbridges and 
of other social phenomena, development of efficient high-frequency power sources, 
modeling and control of neuronal rhythms, etc. 
In this paper we present our view of the recent development of this highly 
interesting field.
\end{quotation}

\section{Introduction}
In the second half of the 17th century Engelbert Kaempfer, Dutch physician, made 
a journey to  South-East Asia and later published a book, describing his trip 
\citep{Kaempfer-1727}.
In particular, in his memoirs he gives an account of a spectacular show,
which happens if a swarm of fireflies occupies a tree: the insects 
``hide their Lights all at once, and a moment after make it appear again with the 
utmost regularity and exactness''. 
This phenomenon of \textit{self-synchronization in a large
population of interacting oscillatory objects} not only remains 
an appealing entertainment --- be it an excursion on a night river in Thailand 
to observe fireflies or cycling in a large group of people, 
equipped with electronic ``bikeflies'', as a part of a festival in Chicago
 --- but it stays in the focus of scientific interest within many decades.
The studies of various aspects of the collective dynamics in large oscillatory networks
attract attention of physicists and applied mathematicians and find applications
ranging from electrochemistry to quantum electronics, and  from bridge engineering 
to social sciences.

A particularly popular and expanding field of applications is neuroscience. 
For the first time the link between synchronous flashing of the fireflies 
and origin of the brain rhythms was established, on a quantitative level, 
by the famous mathematician 
Norbert Wiener in his monograph  on Cybernetics \cite{Wiener-65}, in the chapter 
entitled ``Brain waves and self-organizing systems'', see also Ref.~[\onlinecite{Strogatz-94}]. 
Putting forward the hypothesis 
that the brain waves emerge due to ``phenomenon of the pulling together 
of frequencies'', he questioned, whether the same nonlinear mechanism takes place 
in case of firefiles, crickets, and other species exhibiting collective 
oscillatory behavior,
and suggested that experiments on fireflies and on electronic systems 
can shed light on the brain wave dynamics.

Since that many experiments have been reported, 
including those with electrochemical \cite{Kiss-Zhai-Hudson-02a,Zhai-Kiss-Hudson-08} 
and electronic oscillators \cite{Temirbayev_et_al-12,*Temirbayev_et_al-13},
metronomes \cite{Martens2013}, Josephson junctions~\cite{Benz-Burroughs-91},
laser arrays~\cite{Hirosawa:13},
yeast cells~\cite{Richard-Bakker-Teusink-Van-Dam-Westerhoff-96}, 
and gene-manipulated clocks in bacteria~\cite{Prindle_etal-12}.  
They are complimented by observations of many social phenomena like
pedestrian synchrony on footbridges \cite{Dallard-01a}, synchronous hand clapping in
opera houses \cite{Neda-Ravasz-Brechet-Vicsek-Barabasi-00,*Neda_etal-00},
egg-laying in bird colonies~\cite{Henson_etal-10}, and menstrual synchrony 
\cite{McClintock-71,*Weller-Weller-97,*Arden-Dye-98}
(the latter effect remains controversial~\cite{Ziomkiewicz-06}).
This research was also accompanied by an essential progress in the theoretical description,
that we outline below, along with the open questions.

\section{Solvable models} 
A few years after publication of the Wiener's book, Arthur Winfree presented 
a first mathematical 
description of collective synchrony in a large population of biological 
oscillators \cite{Winfree-67,*Winfree-80}. 
Reducing the dynamics of each oscillator to that of only one 
variable, the phase $\vp$ (we will discuss conditions for this reduction below), 
he proposed the mean-field model
\begin{equation}
\dot\vp_k=\w_k+\frac{\e}{N}\Gamma(\vp_k)\sum_{j=1}^N\ I(\vp_j) \;,
\label{WM}
\end{equation}
where $N\gg 1$ is the number of units, $\w_k$ are their natural frequencies, and $\e$
quantifies the strength of the interaction. The function $\Gamma(\vp_k)$ describes the 
phase sensitivity of the oscillator to an infinitesimal perturbation 
and is typically called the phase response curve, PRC. Notice that the PRC can be 
experimentally obtained by repeating stimulation of an isolated system. 
Finally, the forcing function $I(\vp_j)$ describes the effect of the $j$-th 
unit on the unit $k$. In this setup the coupling is assumed to be global, i.e. 
of the all-to-all type, and functions $\Gamma$, $I$ are assumed to be identical for all 
interactions. 
Thus, the inhomogeneity of the system is due to a distribution of frequencies 
$\w_k$ only. Although the model is quite complicated for the analysis, Winfree has
shown that it exhibits a transition to a macroscopic synchronized state, 
characterized by non-zero mean field $N^{-1}\sum_{j}\ I(\vp_j)$. 
He discovered that the collective synchrony is a threshold phenomenon: 
the transition occurs if the coupling strength $\e$ is large enough or
the inhomogeneity, i.e. the width of the distribution of $\w_k$, 
is sufficiently small. For recent studies on the Winfree model see Refs.
[\onlinecite{PhysRevLett.86.4278,*PhysRevE.75.036218,*PhysRevX.4.011009},%
\onlinecite{Giannuzzi_et_al-07}];
it has been shown that it can be treated analytically at least if
 functions  $\Gamma$, $I$ contain only one Fourier harmonics.

The next pioneering step has been done by Yoshiki Kuramoto in his seminal 
publication 40 years ago \cite{Kuramoto-75,*Kuramoto-84}. 
The model he suggested can be considered as the 
weak-coupling limit of Eq.~(\ref{WM}). Indeed, if $\e\ll \w$, then each 
Eq.~(\ref{WM}) can be averaged over the oscillation period; then each term 
$\Gamma(\vp_k)I(\vp_j)$ yields a function of the phase difference, $g(\vp_j-\vp_k)$, 
so that the averaged model reads  
\begin{equation}
\dot\vp_k=\w_k+\frac{\e}{N}\sum_{j=1}^N\ g(\vp_j-\vp_k) \;.
\label{DM}
\end{equation}
The general case of an arbitrary $2\pi$-periodic function $g$ is discussed later,
while the simplest case $g(\cdot)=\sin(\cdot)$ corresponds to the famous Kuramoto
model (notice that choice of the sine function is not a result of some approximation 
but just the simplest solvable case):
\begin{equation}
\dot\vp_k=\w_k+\frac{\e}{N}\sum_{j=1}^N\ \sin(\vp_j-\vp_k)=\w_k+\e R \sin(\Theta-\vp_k)\;.
\label{KM}
\end{equation}
Here $R\ee^{\ii\Theta}=N^{-1}\sum_j e^{\ii\vp_j}$ is the complex mean field, 
$R$ and $\Theta$ are its amplitude and phase, respectively. 
Kuramoto solved the problem in the thermodynamical limit $N\to\infty$, 
using a self-consistent approach: assuming 
a harmonic mean field with unknown amplitude and frequency, he obtained closed
integral equations for these two quantities.  
The celebrated result is the existence of the 
critical coupling, $\e_c$, proportional to the width of the frequency distribution.
For sub-threshold coupling the mean field is exactly zero, while for $\e>\e_c$, 
a nontrivial solution with non-zero mean field appears; the 
amplitude of the field grows as $\sqrt{\e-\e_c}$ and its frequency
equals the central frequency of the distribution of $\w_k$
(which is assumed to be symmetric and unimodal). 
Thus, appearance 
of the collective mode can be treated as a second-order nonequilibrium 
phase transition.
As has been shown later, the Kuramoto model with the uniform frequency 
distribution exhibits a jump of the order parameter \cite{Pazo-05}.

\section{Collective dynamics of the Kuramoto model}
  
The results of Kuramoto gave an enormous impact on the development of the 
field, with still an increasing number of publications on the subject. 
The model (\ref{KM}) and its extension due to Sakaguchi and Kuramoto 
\cite{Sakaguchi-Kuramoto-86}, who accounted for a possible phase shift in coupling,
\begin{equation}
\dot\vp_k=\w_k+\e R \sin(\Theta-\vp_k+\alpha)\;,
\label{KSM}
\end{equation}
became a paradigmatic model for analysis of large oscillator ensembles.

\subsection{Watanabe-Strogatz theory}

We now briefly discuss an interesting and important mathematical property
of the Kuramoto-Sakaguchi model~(\ref{KSM}), 
namely its \textit{partial integrability}.
We start this discussion by consideration of the simplest case of identical 
units, $\w_k=\w$.
As has been
shown in the seminal publications by Watanabe and 
Strogatz~\cite{Watanabe-Strogatz-93,*Watanabe-Strogatz-94} (WS), the dynamics  of this 
system can be described by three 
global variables $\rho$, $\Phi$, $\Psi$ and $N-3$ constants of 
motion $\psi_k$. Here the variable $0\le\rho\le 1$ is roughly similar 
to the mean field amplitude $R$; $\Phi$ and $\xi$ are angular variables;
often it is convenient to combine two variables as $z=\rho \ee^{\ii \Phi}$, 
see also Refs.~[\onlinecite{Pikovsky-Rosenblum-08,*Pikovsky-Rosenblum-11}]. 
The main idea of the powerful WS theory is as follows. 
Consider $N>3$ identical oscillators governed by 
\begin{equation}
\dot\vp_k=\w(t)+\mbox{Im}\left [H(t)\ee^{-\ii\vp_k}\right ] \;,
\label{WS1}
\end{equation}
where $H(t)$ is an arbitrary common forcing. Obviously, Eq.~(\ref{KSM}) 
is a particular case of Eq.~(\ref{WS1}). The latter can be re-written as 
\begin{equation}
\der{}{t}\left(\ee^{\ii\vp_k}\right)=\ii\w(t) \ee^{\ii\vp_k}+\frac{1}{2}H(t)
		-\frac{\ee^{\ii2\vp_k}}{2}H^*(t)\;.
\label{WS2}
\end{equation}
Next, we transform $N$ variables $\vp_k$ to 
complex $z$, $|z|<1$, and $N$ real $\xi_k$, according to
\begin{equation}	
\label{WS3}
 \ee^{\ii\vp_k}=\frac{z+\ee^{\ii\xi_k}}{1+z^*\ee^{\ii\xi_k}}\;.
\end{equation}
This transformation, found by WS and written in this form
in Refs.~[\onlinecite{Pikovsky-Rosenblum-08,*Pikovsky-Rosenblum-11},%
\onlinecite{Marvel-Mirollo-Strogatz-09}], 
is known as the M\"obius transformation. 
Since the system is over-determined, one requires 
\begin{equation}	
\label{WS4}
 N^{-1}\sum_{k=1}^N \ee^{\ii\xi_k}=\avr{\ee^{\ii\xi_k}}=0\;.
\end{equation}
Substituting Eq.~(\ref{WS3}) into Eq.~(\ref{WS2}) we obtain, after straightforward
manipulations:
\begin{equation}	
\label{WS5}
\begin{split}
\dot{z}&+\left[\dot{z}z^*-z\dot{z}^*
	+\ii\dot{\xi_k}(1-|z|^2)\right]\ee^{\ii\xi_k}
  =\ii\w z+\frac{H}{2}\\
  &-\frac{H^*}{2}z^2
	+\left[\ii\w(1+|z|^2)+(z^*H-zH^*)\right]\ee^{\ii\xi_k}\\
	&+\left[\dot{z}^*+\ii\w z^*+\frac{H}{2}z^{*2}-\frac{H^*}{2}\right]\ee^{2\ii\xi_k}\;.
\end{split}
\end{equation}
Averaging these equations over $k$, using Eq.~(\ref{WS4}) and 
$\avr{\dot\xi_k e^{\ii\xi_k}}=0$, we obtain 
\begin{equation*}	
\dot z-\ii\w z-\frac{H}{2}+\frac{H^*}{2}z^2=
\left[\dot z^*+\ii\w z^*-\frac{H^*}{2}+\frac{H}{2}z^{*2}\right]\avr{\ee^{2\ii\xi_k}}
\end{equation*}
and this equation is obviously satisfied, if  
\begin{equation}
\dot z = \ii\w z+\frac{H}{2}-\frac{H^*}{2}z^2 \;.
\label{WS6}
\end{equation}
Substitution of Eq.~(\ref{WS6}) and its complex conjugate into Eq.~(\ref{WS5})
yields:
\begin{equation*}	
\ii\dot\xi_k(1-|z|^2)=\left [ \ii \w+\frac{z^*H-zH^*}{2}\right ] (1-|z|^2)\;.
\label{WS7}
\end{equation*}
Excluding the fully synchronous case $|z|=1$
and introducing $\xi_k=\alpha+\psi_k$, where all $\psi_k=\const$,
we obtain:
\begin{equation}
\dot\alpha=\w+\mbox{Im}(z^*H)\;.
\label{WS8}
\end{equation}
Expressions~(\ref{WS6},\ref{WS8}) represent the Watanabe-Strogatz equations, which
completely describe the evolution of an ensemble of identical oscillators. 

Extension of the WS theory for the case of non-identical units depends on the 
structure of the ensemble. Consider first the case of a 
\textit{hierarchical population}, which consists of $M$ groups of identical units, 
so that units in each group are subject to the same 
field \cite{Pikovsky-Rosenblum-08}. 
In this case the dynamics of the ensemble obeys the system of $M$ coupled WS 
equations~(\ref{WS6},\ref{WS8}).
Another practically important case is a large population, $N\to\infty$, which can be 
characterized by a distribution of frequencies $g(\w)$. This system is described by WS 
variables $z(\w,t)$, $\alpha(\w,t)$ and the equations for 
$\partial_t{z}$, $\partial_t{\alpha}$ 
are similar to~(\ref{WS6},\ref{WS8}), see Ref.~[\onlinecite{Pikovsky-Rosenblum-11}].

\subsection{From WS to Ott-Antonsen theory}

There exist a particular case, when the WS equations can be 
essentially simplified. 
As has been shown in Refs.~[\onlinecite{Pikovsky-Rosenblum-08,*Pikovsky-Rosenblum-11}], 
for large $N$ and for the uniform 
distribution of constants of motion $\psi_k$, the WS variable $z$ coincides with the 
\textit{local} Kuramoto mean field 
\begin{equation}
Z(\w,t)=\int_0^{2\pi}w(\vp,t|\w)\ee^{\ii \vp}\dd \vp \;,
\label{kurpar}
\end{equation}
(where $w(\vp,t|\w)$ is the distribution density of oscillators' phases 
at frequency $\w$; $\int d\vp \,w(\vp,t|\w)=1$)
to be distinguished from the global mean field
\begin{equation}
Y(t)=\int_{-\infty}^\infty g(\w)Z(\w,t)\dd \vp\;.
\label{globpar}
\end{equation}
As a result, one obtains closed equations 
for $Z$:
\begin{align}
\label{WS9}
\pd{Z(\w,t)}{t} &= \ii\w Z+\frac{H(\w,t)}{2}-\frac{H^*(\w,t)}{2}Z^2 \;, \\[1ex]
\label{WS10}
\pd{\alpha(\w,t)}{t}&=\w+\mbox{Im}\left [ Z^*H(\w,t)\right ] \;.
\end{align}
In the most common case of the mean field coupling $H=\e\ee^{\ii\beta}Y$, 
the forcing $H$ is independent on $\alpha$ and Eq.~(\ref{WS10}) 
becomes irrelevant. Hence, we are left with Eq.~(\ref{WS9}) which also appears
in the recent theory by Ott and Antonsen (OA) \cite{Ott-Antonsen-08,Ott-Antonsen-09}, 
briefly introduced below.

Consider the thermodynamical limit of the model  
\begin{equation}
\dot\vp_k=\w_k+\mbox{Im}\left [H(t)\ee^{-\ii\vp_k}\right ] \;
\label{OA1}
\end{equation}
and the corresponding continuity equation for the distribution of phases 
\begin{equation}
\pd{w}{t}+\pd{}{\vp}(w\dot\vp)=0 \;.
\label{OA2}
\end{equation}
Next, let us introduce Fourier components of the density (the generalized local Daido order parameters)
\begin{equation}
Z_m(\w,t)=\int_0^{2\pi}w(\w,\vp,t)\ee^{\ii m\vp}\dd \vp \;,
\end{equation}
cf. Eq.~(\ref{kurpar}). Computing 
\begin{equation*}
\pd{Z(\w,t)}{t}=\int_0^{2\pi}\pd{w}{t}\ee^{\ii m\vp}\dd \vp =
-\int_0^{2\pi}\pd{(w\dot\vp)}{t}\ee^{\ii m\vp}\dd \vp\;,
\end{equation*}
integrating by parts and using Eq.~(\ref{OA1}), one obtains an 
infinite-dimensional system of ODEs:
\begin{equation}
\pd{Z(\w,t)}{t}=\ii m \w Z_m +\frac{m}{2}\left ( HZ_{m-1}-H^*Z_{m+1}\right )\;.
\label{eq:hierOA}
\end{equation}
A particular case $Z_m=Z_1^m=Z^m$, called the OA manifold, reduces all the 
equations~\eqref{eq:hierOA}
to a
single Eq.~(\ref{WS9}).
Thus, the OA manifold corresponds to the special solution of 
the WS theory, with the uniform distribution of constants of 
motion $\psi$. Furthermore, OA argued that for continuous 
frequency distribution $g(\w)$ the OA manifold is the only attractor 
\cite{Ott-Antonsen-09} (although relaxation to it may be rather 
slow~\cite{Mirollo-12}).
A particular case of the Lorentzian distribution 
$g(\w)=[\pi(\w^2+1)]^{-1}$ admits a further essential simplification.
Under assumption that $Z(\w)$ is analytic in the upper half-plane, 
one computes the integral in Eq.~(\ref{globpar}) by virtue of residues
and obtains $Y=Z(\ii)$. 
Substituting this along with $\w=\ii$ into Eq.~(\ref{WS9}) one obtains
the OA equation
\begin{equation}
\dot Y =\left ( \frac{\e\ee^{\ii\beta}}{2}-1\right )Y-
\frac{\e\ee^{-\ii\beta}}{2}Y^2Y^*
\label{eq:OA}
\end{equation}
for the evolution of the mean field in the Kuramoto-Sakaguchi model.
Looking for a stationary synchronous solution, we set $Y=R_0\ee^{\ii \nu t}$
and obtain the amplitude  $R_0=\sqrt{1-2/\e\cos\beta}$ and frequency 
$\nu=(\e\cos\beta-1)\tan\beta$
of the mean field.

For an illustration of the WS and OA theories we consider the model 
of two interacting populations by Abrams et al.
\cite{Abrams-Mirollo-Strogatz-Wiley-08}. 
Both populations consist of the same number of identical oscillators, 
but the coupling strength within the group differs from that between the groups. 
The dynamics can be fully described by two coupled systems of WS equations, 
and is therefore six-dimensional \cite{Pikovsky-Rosenblum-08,*Pikovsky-Rosenblum-11}. 
In the particular case of uniformly distributed constants $\psi$, i.e. on the OA 
manifold, the dimension reduces to four. The  latter case, studied 
in Ref.~[\onlinecite{Abrams-Mirollo-Strogatz-Wiley-08}], reveals, for certain parameters, 
an interesting solution, when one population synchronizes while the other
stays between complete synchrony and full asynchrony, 
i.e. $R_1=1$ and $0<R_2<1$. 
This symmetry-breaking state is called \textit{chimera}. 
(Notice that originally 
chimera states have been introduced for nonlocally coupled oscillator chains
\cite{Kuramoto-Battogtokh-02,*Abrams-Strogatz-04,*PhysRevLett.100.044105}.)
Interesting, 
that $R_2$ can be time-periodic. Analysis of the full six-dimensional systems
exhibits additional solutions with the quasiperiodic chimera 
states~\cite{Pikovsky-Rosenblum-08,*Pikovsky-Rosenblum-11}.
Theoretical predictions have been confirmed in recent experiments with 
two groups of metronomes, placed on platforms which were coupled via springs
\cite{Martens2013}.

\subsection{Generalizations of the Kuramoto model}
There are many studies of different generalizations of the Kuramoto model. 
Here we briefly
mention those where the coupling function is purely harmonic like in Eq.~\eqref{KSM}, but 
the overall setup is more complex:
\paragraph{Multi-modal frequency distribution and several interacting populations.} We 
have seen that an ensemble with a Lorentzian distribution of frequencies is described by the OA 
Eq.~\eqref{eq:OA}. A multi-modal distribution of frequencies can be modeled as 
a superposition of Lorentzians and, hence, described by a system of 
coupled OA Eqs.~\eqref{eq:OA}, see e.g. Ref.~[\onlinecite{So-Barreto-11}]. Moreover, 
this approach can be generalized to a set of populations with frequencies,
distributed around completely different central values, whereas 
the latter can be either in 
resonance~\cite{Komarov-Pikovsky-13} or not~\cite{Komarov-Pikovsky-11}. 
\paragraph{Nontrivial transitions for unimodal distributions.} 
For a long time it has been assumed that in the Kuramoto-Sakaguchi
model~\eqref{KSM} with different unimodal distributions of frequencies 
the dynamics of the mean field is qualitatively the same as for the Lorentzian 
distribution (Eq.~\eqref{eq:OA}). Rather surprisingly, Omelchenko and Wolfrum
\cite{Omelchenko-Wolfrum-12} have demonstrated rather complex transition scenaria, 
including first-order transitions and bistability, for some unimodal distributions.
\paragraph{Complex coupling schemes.} 
The Kuramoto-Sakaguchi model describes a ``direct'' coupling scheme: 
the mean field, calculated algebraically 
from the states of all oscillators, enters the equations for the phases. 
The coupling scheme can be, 
however, more complex: the mean field may act on some macroscopic variables that 
obey a set of generally nonlinear differential equations, and the acting force is a 
function of these variables. For example,
in  a description of pedestrian synchrony on a footbridge~\cite{Eckhardt_et_al-07}, 
one describes each pedestrian by an individual phase variable,  
but one needs also equations for the swinging mode of the bridge. 
The latter is driven by the field created by all pedestrians 
and, in its turn, affects their gaits. Similarly, 
electronic~\cite{Temirbayev_et_al-12,*Temirbayev_et_al-13} or 
electrochemical~\cite{Kiss-Zhai-Hudson-02a,Zhai-Kiss-Hudson-08} 
oscillators can be coupled through the common 
macroscopic current or voltage, 
which obeys macroscopic equations describing the coupling circuit. 
In this way one also describes 
synchronization of Josephson 
junctions~\cite{Wiesenfeld-Swift-95,*Wiesenfeld-Colet-Strogatz-96} 
or spin-torque~\cite{Grollier-Cros-Fert-06,*Tiberkevich_etal-09,*Pikovsky-13} 
oscillators. 
\paragraph{Nonhomogeneous populations.}
There is another 
generalization of the standard Kuramoto-Sakaguchi coupling. 
The latter assumes that all the oscillators
make the same contribution to the mean field and that the mean field acts 
on all oscillators in an equal way. 
Refs.~[\onlinecite{Iatsenko_etal-14,Vlasov-Macau-Pikovsky-14}] generalized 
this to the situation, where the global field still can be introduced, 
but oscillators contribute to it differently, i.e. with different amplitudes and 
phase shifts, and the field also acts differently on different oscillators. 
For a physical implementation one can can consider
a receiver which collects signals emitted from the oscillators 
(where the attenuation and the phase shift are due to signal propagation properties), 
and the governing signal is then transmitted to the 
oscillators, being also subject to attenuation and phase shift depending on the 
positions of the units~\cite{Vlasov-Macau-Pikovsky-14}. 
One variant of an inhomogeneity of the population is when
it consists of two groups with different reaction to the force~\cite{Hong-Strogatz-11}: some 
are ``conformists'' (they follow the force) and some are ``contrarians'' (they tend to be 
in anti-phase with the forcing). Another possibility, quite general for neural ensembles, 
is when some oscillators are inhibitors (trying to desynchronize others, coupling 
is repulsive) and others exert an excitatory action, attempting to synchronize 
(attracting the phase)~\cite{Zanette-05,Maistrenko-Penkovsky-Rosenblum-14}.
\paragraph{Effects of noise.}
Independent noisy forces acting on oscillators of a population counteract synchronization. 
In this sense noise is a source of disorder, similar to the distribution of natural
frequencies. Due to noise, synchronization is a threshold phenomenon already for 
identical oscillators and the transition occurs at a critical coupling which
is proportional to the noise intensity. In the thermodynamic limit such a 
system is described analytically by a nonlinear Fokker-Planck equation which differs from 
the continuity equation~\eqref{OA2}
by a term $\sim \sigma^2 \frac{\partial^2 w}{\partial\vp^2}$, where $\sigma$ is the 
amplitude of the noise.
Completely opposite is the action of a common noise:
it tends to synchronize the population of oscillators 
and for identical units this effect 
can be described within the WS framework~\cite{Braun-etal-12}. 
\paragraph{Finite-size fluctuations.} Kuramoto and OA theories have been 
developed in the thermodynamic limit of infinitely large populations; 
WS theory applies to any population size, but for identical populations only. 
In finite populations with different natural frequencies of units one expects to
observe fluctuations, both prior and beyond the synchronization transition, 
which is defined ambiguously in this case. 
In Ref.~[\onlinecite{Popovych-Maistrenko-Tass-05}] the Kuramoto model with a uniform 
distribution of frequencies and a relatively small number of oscillators have been 
shown to be chaotic 
prior to synchronization transition, the maximal Lyapunov exponent however decreases 
with the system size as $\lambda\sim N^{-1}$. Above synchronization transition only 
regular dynamics have been observed. 
However, for $N\gg 1$ and close to the synchronization transitions the regime
is complex: if it is not chaotic, then it is quasiperiodic with a large number 
of incommensurate frequencies; here statistical approaches based on finite-size 
scaling have been applied to find critical indices of 
$N$-dependence~\cite{Daido-90,Hong_etal-07}.   
\paragraph{Kuramoto model on networks.} Kuramoto model has been initially formulated 
for the ensemble of globally coupled oscillators. Recently, it has been extensively 
studied for other coupling configurations, i.e. for networks of different complexity, 
including small-world networks~\cite{Hong_etal-02}, multidimensional hypercubic 
lattices~\cite{Hong_etal-07}, networks with a modular 
structure~\cite{Skardal-Restrepo-12}, and an ensemble 
with an extra leading element (hub)~\cite{Kazanovich01122003}. 
Dynamical properties of the transition to synchronization depend on the network 
topology~\cite{Arenas_etal-06}. One of the popular applications here is study
of synchronization of power grids~\cite{PhysRevLett.109.064101}.
\paragraph{External forcing and collective phase resetting.} 
As one can see from Eq.~\eqref{eq:OA}, the macroscopic order parameter obeys an 
equation for a self-sustained oscillator. Thus, the collective
mode has the same properties as such an oscillator. In particular, if the ensemble is
forced by a periodic force, the latter can entrain the frequency of the mean field 
oscillations~\cite{Sakaguchi-88,*Childs-Strogatz-08,*Baibolatov2009,Ott-Antonsen-08}. 
In the case of a pulse force, the latter can shift the phase of the collective mode; 
in this way one defines the phase response curve for the 
collective mode~\cite{Kawamura_etal-08,*Ko-Ermentrout-09,*Levnajic-Pikovsky-10}. 
\paragraph{Mathematical results.} The Kuramoto model has been extensively studied 
on the physical and computational level,
but rigorous mathematical results are sparse. They mainly refer to stability analysis 
of the desynchronized state~\cite{Mirollo-12,Strogatz-Mirollo-91,*Fernandez_etal-14}; 
for a description of the synchronization transition as a bifurcation problem see 
Refs.~[\onlinecite{Chiba-Nishikawa-11,*Chiba-13,*Dietert-14}].

\section{General coupling functions}
Now we come back to Eq.~(\ref{DM}). For a general coupling function $g$, 
this equation represents the Daido model \cite{Daido-93,*Daido-93a,*Daido-96,Daido-95}.
Expanding $g$ into Fourier series, $g(\eta)=\sum_n g_n \ee^{\ii n\eta}$, 
and introducing generalized order parameters 
\begin{equation}
Z_n=N^{-1}\sum_{j=1}^N\ee^{\ii n \vp_j}\;,
\label{DM1}
\end{equation}
one can re-write the model as	
\begin{equation}
\dot\vp_k=\w_k + \e \sum_{n}g_nZ_n\ee^{-\ii n \vp_j}\;.
\label{DM2}
\end{equation}
One can see that generally all order parameters should be determined self-consistently, 
so that a complete analysis at general coupling is still missing. We mention 
here several interesting 
regimes appearing due to high harmonics in coupling function.
\paragraph{Clustering.} Even for identical oscillators the WS theory does not apply, 
and a population can build several clusters~\cite{Okuda-93}, 
each of them consisting of fully synchronized units.
\paragraph{Heteroclinic cycles.} In Ref.~[\onlinecite{Hansel-93}] nontrivial regimes of 
clustering and switching between different cluster states have been found in a
population of identical oscillators with a function $g$ containing the 
first and the second harmonics. 
This complex dynamics is due to 
a heteroclinic cycle in the phase space, well understood for small 
networks~\cite{Ashwin_etal-07,*Karabacak-Ashwin-10}.
\paragraph{Multi-branch entrainment.} Already for two harmonic components 
in the coupling function $g$, the r.h.s. of Eq.~\eqref{DM2} as a 
function of $\vp$ can have two stable branches.
This is a new situation compared to the standard Kuramoto model: 
now for a given mean field entrainment 
at two different microscopic phases is possible. 
This leads to an enormous multiplicity of microscopic states and
to a complex structure of macroscopic 
regimes~\cite{Daido-95,Daido-96a,Komarov-Pikovsky-13a,*Komarov-Pikovsky-14}. 

\section{Non-phase models}
We started our discussion of collective ensemble dynamics from phase models.
This approach relies on the well-known idea that motion along the limit 
cycle of an autonomous system can be parameterized by a single variable, 
the phase. Moreover, if the interaction of the oscillator with the 
environment is weak so that one can neglect the deviation of the trajectory
from the cycle of the autonomous system, then the low-dimensional phase 
description remains valid. 
Generally, for strong coupling, one has to analyze full dynamical models which is a complicated
problem that can hardly be treated analytically.

Numerical studies reveal that transition to collective synchrony is a quite
general property, observed for various periodic, noisy, 
and even chaotic oscillators \cite{Pikovsky-Rosenblum-Kurths-96,Kiss-Zhai-Hudson-02a}, 
including, e.g., spiking and bursting model neurons.
The picture is, however, non-universal. The most transparent and studied model is 
ensemble of coupled Stuart-Landau 
oscillators~\cite{Nakagawa-Kuramoto-93,*Nakagawa-Kuramoto-94} (this oscillator is the simplest prototype of a limit-cycle oscillator, with a perfect separation of amplitude and phase variables,
see Eq.~\eqref{nlSL} below), 
and essentially new effects are the
oscillation quenching, when too strong coupling effectively introduces 
additional damping to ensemble elements, and the collective chaos in a system
of initially periodic units. 

On the other hand, non-identical chaotic 
phase-coherent R\"ossler oscillators adjust their frequencies and phases
(this effect is known as phase synchronization of chaos 
\cite{Rosenblum-Pikovsky-Kurths-96})
and produce a nearly periodic mean field. The oscillators themselves remain 
chaotic, but irregular fluctuations of the amplitudes turn to be averaged out 
in the mean field; the small fluctuations of the latter are presumably due 
to the finite-size effect \cite{Pikovsky-Rosenblum-Kurths-96}.
Experimental studies on chaotic electrochemical oscillators 
\cite{Kiss-Zhai-Hudson-02a} confirm theoretical predictions.

Another important class are rotators: these systems are described by an angle-like variable, 
which is very similar to the phase, and do not have amplitudes. If an 
equation governing the rotator's angle is one-dimensional, the dynamics can be 
reduced to a Kuramoto-type phase model, what has been extensively discussed in context
of Josephson junctions~\cite{Wiesenfeld-Swift-95,*Wiesenfeld-Colet-Strogatz-96}. 
Quite often the inertia of rotators cannot be neglected and, hence, they are described by 
a second-order equation for the angle variable. 
In this case the dynamics can strongly deviate from that of the Kuramoto model, 
e.g. transition to synchrony may exhibit hysteresis, similarly to a first-order 
phase transition~\cite{Tanaka-Lichtenberg-Oishi-97}.
A particular subclass is constituted by globally coupled rotators without damping: 
this conservative system yields the so-called Hamiltonian Mean Field (HMF) 
model~\cite{Antoni-Ruffo-95}; 
see Refs.~[\onlinecite{Campa200957,*Gupta_etal-14}] for a review of recent results and 
relation to the Kuramoto model.

\section{Complex collective dynamics around synchrony}
We have discussed in detail the main effect observed in globally coupled systems,
namely emergence of the collective mode, which is well-understood in the simplest case,
when many units synchronize and therefore there outputs sum up coherently, 
contributing to this mode. We have also mentioned, that the mode itself can exhibit
chaotic dynamics. Now we discuss other, less explored, situations, when the collective 
dynamics of an ensemble is complex.

\subsection{Partial synchrony}

In case of the Kuramoto model of identical oscillators, only two situations are possible:
full synchrony for attractive coupling (order parameter equal to one) and fully 
asynchronous state (zero order parameter) for repulsive coupling.  However, this situation 
is not general and we can face a case, when both fully synchronous and fully asynchronous 
states are unstable, so that the system settles somewhere in between, at a state which 
is often called partial synchrony. The simplest and well-known 
example of partial synchrony is 
clustering; below we discuss several other situations where oscillators are not
organized in clusters. 

We examplify partial synchrony with $N\gg 1$ mean field 
coupled oscillators \cite{Ehrich-Pikovsky-Rosenblum-13}:
\begin{eqnarray}
\dot x_k&=& y_k-x_k^3+3 x_k^2-z_k+5+\e(X-x_k) \;,\nonumber\\
\dot y_k&=& 1-5x_k^2-y_k  \;,\label{hmr}\\
\dot z_k&=& 0.006 [4 (x_k+1.56)-z_k] \nonumber \;,
\end{eqnarray}
where $k=1,\ldots,N$ and $X=N^{-1}\sum_{j=1}^N x_j$. 
Here individual units represent a popular Hindmarsh-Rose neuronal model \cite{Hindmarsh84}.
The quite standard parameter values used here correspond to a limit-cycle solution for the 
uncoupled neurons, or, in neuroscience language,  
to a state of periodic spiking. Fully synchronous state, $x_1=x_2=\ldots=x_N$, 
$y_1=y_2=\ldots=y_N$, $z_1=z_2=\ldots=z_N$ is obviously a solution of the system. 
However, stability of this state depends on the coupling strength $\e$, as can be checked 
numerically by virtue of computation of the transversal stability via evaporation multipliers 
\cite{Kaneko-94,*Pikovsky-Popovych-Maistrenko-01,Rosenblum-Pikovsky-07}. 
This analysis, confirmed by direct simulation, 
demonstrates, that there exist a critical coupling value $\e_c$, at which
the synchronous solution loses its stability via a Hopf-like bifurcation, i.e. two 
complex multipliers cross the unit circle, giving birth to a new frequency. 
(Notice that the stability of the synchronous solution is re-gained for very large $\e$.)
Beyond the synchrony breaking, the states of oscillators 
in the phase space form a thin stripe, stretched along 
the limit cycle; the points within this stripe slowly interchange their position,
with a characteristic time of tens of periods. 
Roughly speaking, the average frequency of all oscillators and of the mean field is the same, but 
the phase shift of the oscillators with respect to the field is slowly modulated.
As a result, the dynamics looks quite complicated and 
is possibly weakly chaotic~\cite{Ehrich-Pikovsky-Rosenblum-13}.

For another example we consider a popular model 
of a series array of resistively shunted Josephson junctions. 
The junctions are coupled by virtue of a parallel 
$RLC$-load~\cite{Wiesenfeld-Swift-95,*Wiesenfeld-Colet-Strogatz-96}. 
As has been shown in cited refs., for a weak coupling and linear load,
the system 
is equivalent to the Kuramoto model. Consider now a \textit{nonlinear coupling}; namely, 
let the inductance in the $RLC$-circuit be nonlinear so that 
the magnetic flux depends on the current $\dot Q$
through the $RLC$-load as $\Phi=L_0\dot Q +L_1 \dot Q^3$. 
Numerical study~\cite{Rosenblum-Pikovsky-07}
reveals a synchrony-breaking transition: 
at $\e=\e_c$ the synchronous state becomes unstable; at this critical coupling value real 
evaporation multiplier $\mu$ becomes larger than one (in contradistinction
to example~\eqref{hmr} where the multiplier is complex). 
For $\e>\e_c$ the systems is in a state 
of partial synchrony, with the order parameter being a smooth decreasing function of $\e$.
Furthermore, the dynamics becomes quasiperiodic: the frequency of the mean field is larger than 
the frequency of individual junctions, so that the latter are not locked to the field.
The frequency difference grows with $\e-\e_c$.
 
A general theory of partially synchronous states which appear after the synchrony breaking 
is still missing and requires further investigations. Below we present an analytically 
tractable and relatively transparent example.

\subsection{Self-organized quasiperiodic dynamics}
Now we analyze the system of $N\gg 1$ Stuart-Landau oscillators:
\begin{align}
\label{nlSL}
\dot a_k=(1+\ii \w)a_k-(1+\ii\alpha)|a_k|^2a_k+(\xi_1+\ii\xi_2)A\\ \nonumber
-(\eta_1+\ii\eta_2)|A|^2A\;,
\end{align}
where $A=N^{-1}\sum_{j=1}^N a_j$ and $\xi_{1,2}$, $\eta_{1,2}$ are coupling parameters. 
This model differs from that of Refs.~[\onlinecite{Nakagawa-Kuramoto-93,*Nakagawa-Kuramoto-94}]
due to the nonlinearity in coupling. In the weak coupling approximation the 
model with purely linear coupling~\cite{Nakagawa-Kuramoto-93,*Nakagawa-Kuramoto-94} (i.e. with
$\eta_1=\eta_2=0$) yields the 
Kuramoto-Sakaguchi Eq.~(\ref{KSM}), while the nonlinear Eq.~(\ref{nlSL})
reduces to a particular case of the following phase model 
\cite{Pikovsky-Rosenblum-09}:
\begin{equation}
\dot \vp=\w + \e \alpha(\e,R)\sin[\Theta-\vp_k+\beta(\e,R)] \;
\label{nlKS}
\end{equation}
Equation (\ref{nlKS}) can be considered as an extension of the Kuramoto-Sakaguchi model.
Here $R$ is the mean field amplitude and the bifurcation parameter $\e$ 
corresponds to a combination of the parameters $\xi_{1,2}$, $\eta_{1,2}$ 
and $\alpha(\e,R)$, $\beta(\e,R)$ are some functions. Notice that Eq.~(\ref{nlKS}) 
appears also in a model of Stuart-Landau oscillators coupled via a common
nonlinear medium \cite{Rosenblum-Pikovsky-07,Pikovsky-Rosenblum-09}, similarly 
to the Josephson junction model.
 	
Let us consider the effect of these functions separately, starting with the 
case when $\beta=\const$, $|\beta|<\pi/2$, i.e. the coupling is attractive,
cf. Refs.~[\onlinecite{Filatrella-Pedersen-Wiesenfeld-07},\onlinecite{Giannuzzi_et_al-07}].	
Suppose that $\alpha$ is a decreasing function 
of $\e$, e.g. $\alpha(\e,R)=(1-\e R^2)R$ 
(this function indeed appears for a certain combination of parameters in 
Eq.~(\ref{nlSL})). For $\e<\e_c=1$ this function is positive for fully 
synchronous case $R=1$ and, hence, this state is stable. For $\e>\e_c$
we have $\alpha(\e,1)<0$, i.e. the coupling becomes repulsive. As a result, the 
system stays at the border between attraction and repulsion, determined
by the condition $\e R^2=1$, forming a \textit{self-organized bunch state}
\cite{Pikovsky-Rosenblum-09}. In this state, for general initial conditions 
the oscillator phases spread around the unit circle so that $R=1/\sqrt{\e}$;
this bunch is stationary in the coordinate frame, rotating with the frequency 
$\w$.
 
Now we analyze a more interesting case when $\alpha=R$ and 
$\beta(\e,R)=\beta_0+\beta_1\e^2 R^2$, $|\beta_0|<\pi/2$, $\beta_1>0$.
It is easy to see that synchrony ($R=1$) is stable if
$\beta_0+\beta_1\e^2<\pi/2$ and becomes unstable when $\e$ exceeds the 
critical value $\e_c=\sqrt{(\pi/2-\beta_0)/\beta_1}$.
Again, the system settles at the border of stability, so that the 
condition $\beta_0+\beta_1\e^2 R^2=\pi/2$ is fulfilled. 
However, in this case the dynamics exhibit a peculiar feature, also
possessed by the Josephson junction model discussed above:
the frequency of the mean field differs from the frequency of the 
individual units. Thus, the state can be characterized by two generally 
irrationally related frequencies, and therefore is denoted
as  self-organized quasiperiodicity (SOQ). 
Qualitatively, the emergence of quasiperiodic motion can be 
explained as follows. For $\e>\e_c$ the system is partially synchronous,
i.e. $R<1$ and all oscillators have different phases (notice 
that for general	initial conditions without clusters the phases 
must be different according to Eq.~(\ref{WS3})). Hence, the instantaneous 
frequencies of oscillators differ and, therefore a stationary
(in a rotating coordinate frame) distribution is not a solution.
Quantitative analysis of SOQ dynamics with computation of the 
frequencies of the collective mode and of the oscillators 
can be found in Refs.~\cite{Rosenblum-Pikovsky-07,Pikovsky-Rosenblum-09}.
SOQ states in real-world systems have been demonstrated
in experiments with electronic circuits with global nonlinear 
coupling~\cite{Temirbayev_et_al-12,*Temirbayev_et_al-13}.

To complete the discussion of this issue we mention that similar
complex states with a nonzero collective mode can appear also without
desynchronization transition. For some systems, the fully synchronous
and completely asynchronous states are unstable already for infinitely 
small coupling. A well-studied example is the van Vreeswijk model
of globally coupled leaky integrate-and-fire neurons
\cite{vanVreeswijk-96,*Mohanty-Politi-06}. 
Another example is the emergence of dephasing and bursting
in a system of Morris-Lecar neuronal models 
\cite{Han-Kurrer-Kuramoto-95}; 
computation of the evaporation multipliers for this system shows that
the synchronous state is unstable for the positive coupling range, 
where the effect is observed.
	
\subsection{Chimeralike states in globally coupled systems}

We have already mentioned a symmetry-breaking chimera state in a
system of two interacting Kuramoto populations. Now we discuss emergence
of a chimeralike state in a single homogeneous population.
At first glance, such regimes are not allowed, because identical units
subject to a common force should exhibit the same dynamics 
(i.e. to be either all 
synchronized or all desynchronized). On the other hand, 
there is a number of numerical observations of the mixed states, where only 
a fraction of the population merges to one or several clusters, 
while other elements remain scattered \cite{Kaneko-90b,*Daido-Nakanishi-06}, 
see also recent studies of chimera states in linearly and nonlinearly 
coupled Stuart-Landau oscillators \cite{Sethia-Sen-14,*Schmidt_etal-14}. 
The conditions for emergence of such mixed states 
are not yet clear.
Nevertheless, we can outline one mechanism which results in splitting 
of the population into coherent and incoherent groups. 

Identical elements subject to the same force can behave differently if
they are bistable (multistable), i.e. possess at least two nontrivial 
attractors. 
Another requirement is that the mean field forcing shall be synchronizing 
for the units on one attractor and repulsive for the elements on the 
other one. Taking into account that this partially coherent and partially 
incoherent state shall be maintained self-consistently, we conclude that 
this condition is not trivial. To illustrate this mechanism, we first 
consider a rather artificial but transparent 
example~\cite{Yeldesbay-Pikovsky-Rosenblum-14}, where the oscillators
are the non-isochronous modified Stuart-Landau systems. The modification 
refers to the nonlinearity: in addition to the 3rd order term $|a_k|^2a_k$
we add also the terms $\sim|a_k|^4a_k$,  $\sim|a_k|^6a_k$, cf. Eq.~(\ref{nlSL}). 
As a results the systems possess two stable limit cycles with different 
frequencies, $\W_2>\W_1$. 
The global coupling has its 
own dynamics, cf. 
Refs.~[\onlinecite{Wiesenfeld-Swift-95},\onlinecite{Strogatz_et_al-05}], 
so that the 
mean field forces a harmonic oscillator, 
$\ddot u +\gamma \dot u +\eta^2 u =N^{-1}\sum_j^N \mbox{Re}(a_j)$,
and its output $\dot u$ acts on the  Stuart-Landau systems.
Suppose now that parameters are chosen in such a way that 
$\W_1<\eta<\W_2$. Since the phase shift introduced by the harmonic
oscillator in the last equation is frequency-dependent, the 
coupling synchronizes the large amplitude limit-cycle oscillations,
but prevents synchronization of the low-amplitude ones. 

Much less trivial is emergence of coherent-incoherent states 
in an ensemble of globally coupled 
oscillators with internal delayed feedback loop. 
Such oscillators 
naturally appear, e.g. in laser optics \cite{Masoller-02} as well as
in numerous biological applications 
\cite{Glass-Mackey-88,*PhysRevLett.85.2026,*Batzel-Kappel-11}.
In the simplest case the autonomous dynamics of an oscillator 
with a delayed feedback loop can be  described by a phase model,
$\dot\vp=\w+\alpha \sin(\vp_\tau-\vp)$, 
where $\vp_\tau\equiv\vp(t-\tau)$, $\tau$ is the delay, and
$\alpha$ quantifies the feedback 
strength~\cite{Masoller-02,Niebur-Schuster-Kammen-91,*Stphenyeung-Strogatz-99,%
*Goldobin-Rosenblum-Pikovsky-03}. 
Assuming the global coupling in the ensemble of such oscillators to 
be of the Kuramoto-Sakaguchi type, one writes the model as
\cite{Yeldesbay-Pikovsky-Rosenblum-14}:
\begin{equation}
\dot\vp_k=\w+\alpha\sin(\vp_{\tau,k}-\vp_k) +
\e R\sin(\Theta-\vp_k+\beta)\;.
\label{eq:ph}
\end{equation}
Stability analysis of the fully synchronous one-cluster state yields that
it is unstable for $\beta>\pi/2$. However, numerical simulation shows that 
for $\beta \gtrsim \pi/2$ the asynchronous state with zero mean field, $R=0$,
is also unstable (see \cite{Yeldesbay-Pikovsky-Rosenblum-14} for other parameter
values). Thus, the system ``chooses'' a state between full coherence and 
full incoherence, and in a range of parameters this state is reminiscent of 
a chimera: there is one big cluster (about 70\% of the population size) and 
a cloud of asynchronous oscillators. The frequency of the cluster is larger than 
the frequency of the cloud oscillators. Noteworthy, these states appear
in the case when autonomous oscillators are monostable, though close to the 
bistability domain (for sufficiently strong feedback, 
$\alpha\tau>1$, one time-delayed unit admits multiple periodic solutions). 
It means that the systems become bistable due to coupling and, 
thus, the chimeralike state emerges due to the dynamically sustained bistability.

\section{An important application: neuroscience}
Within fifty years since Norbert Wiener hypothesized the role
of the collective synchrony in the brain dynamics, importance
of this phenomenon for neuroscience became highly recognized. 
It is now well-accepted that macroscopic neural rhythms appear 
due to coordination of firing and/or bursting of interconnected 
neuronal network and the Kuramoto model, in spite of its 
simplicity, became a paradigmatic and widely used setup in
this field, see~\cite{Breakspear-Heitmann-Daffertshofer-10} for a review.
Noteworthy, the neuroscience applications turned out to be 
very fruitful for the theoretical development, 
posing quite interesting, from the viewpoint of nonlinear dynamics, 
problems. In particular, in the context of complex dynamical states 
which are of general interest, we have several times mentioned neuronal 
models, e.g. the Hindmarsh-Rose one. Below, without an ambition to 
provide a comprehensive review, we outline several relevant issues.  

For neuronal models, synchronization in a fully-connected network 
was observed for map-based and continuous time models, both for the 
regimes of periodic spiking and bursting, see e.g. 
Ref.~\cite{Rosenblum-Pikovsky-04a}. 
A bit more detailed models consist of excitatory and inhibitory 
neuronal subpopulations~\cite{Rosenblum-Tukhlina-Pikovsky-Cimponeriu-06}. 
Another issue is a detailed description of synaptic coupling 
\cite{Rulkov-Timofeev-Bazhenov-04}, also with account of 
plasticity \cite{PhysRevE.89.062701,*Popovych-Yanchuk-Tass-13}. 

Fully connected network is certainly a rather crude approximation for 
a neuronal population. However, in some cases it is quite reasonable, as
indicated by a numerical study of randomly coupled networks of map-based 
neurons \cite{Rosenblum-Tukhlina-Pikovsky-Cimponeriu-06}: 
if every unit is connected only to 0.5\% of elements in the ensemble, 
then synchronization properties are practically indistinguishable 
from the fully connected case. For many problems the assumption of 
random connectivity is not appropriate; in this case one has to use
physiologically motivated connectivity structure.  An interesting approach has been 
recently suggested to treat large random networks of coupled oscillators~\cite{Burioni_etl-14}.
To adequately perform the thermodynamic limit and preserve disorder due to randomness
of connections, a heterogeneous mean-field approach has been developed
in which disorder remains the same while the size of the system grows. 
This approach yields a description of  
both microscopic and global features of neuronal synchrony in the model~\cite{Burioni_etl-14}.
Another interesting observation relates to diluted random networks of spike-coupled 
neurons~\cite{Zillmer_etal-06}: while for weak coupling synchrony establishes quite quickly, 
for large coupling
a very long (in fact, exponential in the network size)
transient disordered state is observed, characterized by a negative largest Lyapunov
exponent (so-called ``stable chaos'').


\subsection{Control of collective synchrony}
As an example of a particular application of the discussed theoretical 
ideas we address the problem of suppression of the collective synchrony. 
This problem is relevant for a medical technique, called deep 
brain stimulation (DBS). This technique is exploited to treat the
Parkinson's disease if it cannot be cured by medication. The DBS implies 
electrical stimulation of deep brain structures through the implanted 
micro-electrodes. The modern devices use the constant frequency 
(ca. $100-120$Hz) pulse stimulation, which is typically applied around the 
clock. Although the exact mechanisms of DBS are not yet quite understood,
it is widely used to reduce the limb tremor.
	
Analysis of electrical or magnetic brain activity shows that Parkinsonian 
tremor is associated with the pronounced spectral power at 10 to 12 Hz
\cite{Tass_et_al-98}. Hence, it is reasonable to hypothesize that this
pathological rhythm emerges due to synchronization 
in a neuronal population and to consider the DBS as a desynchronization 
problem \cite{Tass-99,*Tass-2003,*Lysyansky-2011}.
The main idea is then to develop efficient techniques
which reliably suppress the unwanted activity with minimal stimulation.  
There are several directions in these studies. The first one implies
open-loop stimulation with specially organized pulses~\cite{Lysyansky_etall-11} 
through several 
closely spaced cites; the main assumption is that these pulses 
cause formation of several symmetrically positioned clusters, so that 
the mean field vanishes.  

Another direction in  
control of collective oscillation, based on the closed-loop feedback,  was suggested
in~Refs.~[\onlinecite{Rosenblum-Pikovsky-04},\onlinecite{Rosenblum-Pikovsky-04a}]
and verified in 
experiments with electrochemical oscillators in~Ref.~[\onlinecite{Zhai-Kiss-Hudson-08}].
The idea is quite transparent. 
Consider a globally coupled system: all elements are forced
by the collective mode and synchronize due to this forcing. Suppose
we measure the collective oscillation and feed it back, shifting its phase 
and properly amplifying it so that the feedback signal exactly 
compensates the mean field. In this case the oscillators become unforced
and naturally desynchronize due to frequencies mismatch, internal noise, etc. 
Since the units desynchronize, the mean field tends to a constant, 
and so does the feedback signal. The constant component of the feedback 
signal can be easily eliminated; in this way one performs a 
\textit{vanishing stimulation} control. 
It means that stimulation tends to zero as soon as the undesired 
oscillation is suppressed; this property is
extremely important for medical applications. 
The vanishing stimulation control can be implemented via a delayed feedback 
\cite{Rosenblum-Pikovsky-04a,Rosenblum-Tukhlina-Pikovsky-Cimponeriu-06} 
(see Ref.~[\onlinecite{Popovych_etal-05}] for a variant with nonlinear delayed 
feedback which is however not vanishing)
or via a phase shifting passive system without delay 
\cite{Tukhlina-Rosenblum-Pikovsky-Kurths-07}. Moreover, 
an adaptive  strategy can be implemented \cite{Montaseri_et_al-13}, 
so that the desired state can be achieved in spite of unknown parameters 
of the system to be controlled.
Notice that adjusting the phase shift introduced by the feedback loop one 
can, depending on the goal, suppress or enhance the collective synchrony.
	
\section{Surprises and Outlook}
In the first years after development of ideas of synchronization in large ensembles,
especially after Y. Kuramoto constructed his self-consistent theory, it looked like
the very phenomenon of synchronization transition is rather simple and universal.
In this paper we tried to show how in fact nontrivial is even the simplest setup:
features like partial integrability, existence of an exact low-dimensional
manifold, nontrivial transition scenaria even for unimodal distributions of frequencies,
chimera states, self-organized quasiperiodicity occur already in the simplest case of 
sine-coupled phase oscillators. 
For generalizations of this model one observes a plethora of dynamical phenomena 
which is still far from being exhausted. 
Furthermore, addition of complexity to the basic model, e.g. by  
consideration of coupled oscillators on networks, 
results in further growth of the diversity of possible regimes. 

It seems to us that in the nearest future we will experience spreading of the
synchronization theory far beyond its initial scope of nonlinear dissipative dynamical 
systems, e.g. to cover quantum objects~\cite{PhysRevLett.106.257401,*PhysRevLett.113.154101}. 
On the experimental level,
advanced methods of oscillators' control and of data analysis will possibly reveal 
microscopic details of nontrivial synchronization patterns. 
On the other hand, growing interest of mathematicians to these problems 
indicates that the field may become a part of mathematical physics as well.

\end{document}